\documentclass[twocolumn,pre,showpacs,floatfix]{revtex4}
\usepackage{bm}
\usepackage{array}
\usepackage{color}

\newcommand{\bdchange}[1]{{#1}}

\newcommand{\gtapprox}{{\raise.3ex\hbox{$>$\kern-.75em\lower1ex\hbox{$\sim$}}}}
\newcommand{\keff}{k_{eff}}
\newcommand{\lf}{\ell_f}
\newcommand{\bfx}{{\bf x}}
\newcommand{\kdist}{P_K}
\newcommand{\shrmod}{K}
\newcommand{\strmod}{\bar{K}}

\usepackage[dvips]{epsfig}

\usepackage{float}
\usepackage{amsmath}
\usepackage{amssymb}
\usepackage{amsfonts}

\begin{document}
\author{B.A. DiDonna}
\affiliation{Institute for Mathematics and its Applications,
University of Minnesota, Minneapolis, MN 55455-0436, USA}
\author{Alex J. Levine}
\affiliation{
Department of Chemistry and Biochemistry\\ and\\ The California Nanosystems Institute\\
University of California,
Los Angeles, CA 90095, USA}

\title{\bdchange{Unfolding} cross-linkers as rheology regulators in F-actin networks}

\date{\today}

\begin{abstract}
We report on the nonlinear mechanical properties of a statistically
homogeneous, isotropic semiflexible network cross-linked by polymers
containing numerous small unfolding
domains, \bdchange{such as the ubiquitous F-actin cross-linker Filamin.}
 We show that the inclusion of such proteins
has a dramatic effect on the large strain behavior of the network.
Beyond a strain threshold, which depends on network density, the unfolding of
protein domains leads to bulk shear softening. Past this critical strain, the
network spontaneously organizes itself so that an appreciable fraction
of the Filamin cross-linkers are at the threshold of domain
unfolding. We discuss via a simple mean-field
model the cause of this network
organization and suggest that it may be the source of power-law relaxation
observed in {\em in vitro} and in
intracellular microrheology experiments.
We present data which fully justifies our model for a simplified
network architecture.
\end{abstract}


\maketitle

\section{\label{sec:intro}Introduction}

The cytoskeleton of eukaryotic cells can be described as a biopolymer
gel or cross-linked network\cite{alberts:88,pollard:86,elson:88}.  The
principal constituent of the cytoskeleton is a stiff protein
aggregate, F-actin, that is cross-linked densely on the scale of its
own thermal persistence length. In this sense the mesoscale structure
of this biopolymer network differs substantially from that of
synthetic polymer gels\cite{rubenstein:03}, which may be thought of as
interconnected random walks.  This realization of the profound
differences in mesoscale structure between the cytoskeleton, a
prototypical semiflexible filament network, and flexible polymer gels
calls into question the application of standard rubber elasticity
theory to describe the mechanics of semiflexible networks.

A number of researchers have begun to explore the quantitative
relation between the novel microstructure of semiflexible gels and
their observed mechanical
properties~\cite{janmey:90,kroy:96,satcher:96}.  More recently, it has
been proposed that, due to their different architecture, these {\em
semiflexible gels} have macroscopic linear moduli that are generically
more sensitive to their chemical composition
\cite{head:03a,wilhelm:03,head:03b,head:03c,levine:04,didonna:06} than
traditional flexible gels.  Their response to point forces over
mesoscopic distances is much more complex than suggested by the
predictions of continuum elasticity\cite{head:05}. In addition they
exhibit a highly tunable (through network microstructure) nonlinear
response to stress\cite{gardel:04}.

Understanding the biophysical properties of
the cytoskeleton demands a closer examination
not only of the material properties of
generic semiflexible networks, but also
of the chemical heterogeneity
of the physiological cytoskeleton. Until recently, numerical and theoretical
modeling of the  relationship between the
network architecture on the mesoscale and the long-length scale
mechanical properties of these gels
has focused exclusively
on semiflexibility of the filaments~\cite{onck:05}
while ignoring the details
of the chemical agents which cross-link them.
Recent \emph{in vitro} experiments, however, have shown that the
nanoscale mechanics of the physiologically ubiquitous cross-linking proteins
have a significant effect on the mechanics of cross-linked F-actin networks~\cite{gardel:06}.
One particularly interesting class of cross-linkers from the standpoint of
network mechanics are those that contain
unfolding domains such as $\alpha$-Actinin\cite{labeit:95,reif:97} and
Filamin\cite{schwaiger:04,brockwell:05}. Such cross-linkers have
protein domains along their backbone that unfold reversibly at a
critical pulling force. It is still a matter of debate what the
function of these domains may be, but one may speculate that, by
exposing new chemical groups in the unfolded domains, cross-linking
agents such as Filamin may play a role in transducing local
network strain into biochemical signals.

In this paper we investigate the purely mechanical effect of
cross-linkers which have unfolding domains. The
mechanical effect of domain unfolding occurs only at some
finite applied stress, so the effect we wish to study is evident only
in the nonlinear elastic response of the material observed under
finite strain conditions. In this aspect the present work differs from
much of the recent theoretical research on semiflexible network
mechanics that focused on the linear response regime.

We demonstrate two effects that the
inclusion of unfolding cross-linkers
has on the elastic properties of a otherwise
generic semiflexible network. Firstly, since the cross-linkers
can only sustain a finite maximum tensile stress before unfolding,
their inclusion leads to shear softening of the network
at large strain.
Secondly, past the onset of shear softening the network spontaneously
organizes so that the population of
cross-links at given tension grows exponentially or faster
up to the unfolding
force of the domains. Thus at moderate applied stresses the system
appears to adjust its mechanical properties so as to achieve a strain
state in which a significant fraction of its cross-linkers are poised
at the unbinding transition of their internal domains. We refer to such cross-linkers
as being in a critical state. In a thermalized
system, the rapidly rising population of cross-links weighted towards
the unfolding force would yield a fragile state of the material characterized by a
broad, continuous
distribution of relaxation time-scales via thermally excited
subcritical cross-link unfolding. This thermally excited unfolding may account
for the broad, \emph{i.e.} power law distribution of stress relaxation times that has
been observed in {\em in vivo}
experiments\cite{hoffman:05,fredberg:06,fabry:01, laudadio:05, bursac:05} and
has been termed soft-glassy rheology
(SGR)~\cite{sollich:98}.
While in our purely mechanical model this fragile state, in which critical cross-linkers
predominate,
comes about as a consequence of
applied stress, an {\em in vivo} network built with \bdchange{unfolding} cross-linkers
may generically evolve into
this high susceptibility state under the action of internal molecular
motors ({\em e.g.} Myosin -- not considered in our model).
The necessity of applied stress to enter this fragile state in absence of molecular motors
appears to be consistent with the recent finding that pre-stress is necessary
to replicate \emph{in vivo} rheology measurements in artificially synthesized
Actin-Filamin networks~\cite{gardel:06}.

The remainder of this article is arranged as follows.  In section
\ref{sec:theory} we present a simple mean-field theory for the
mechanical effects of cross-linkers with unfolding domains.  In
section \ref{sec:numerics} we discuss  our numerical model
used to study the strain response of an
F-actin network \bdchange{with unfolding cross-links}.
\bdchange{Sections~\ref{sec:theory} and~\ref{sec:numerics}  reiterate
material presented in our initial article on this subject~\cite{didonna:06prl},
while the following sections go beyond the preliminary
analysis presented therein.}
We present the results of this
modeling in section \ref{sec:simresults} and compare them to our
theory. Finally we conclude in section \ref{sec:discussion} where we
discuss the generality of our results and place them in the broader
context of the mechanics of semiflexible gels and the
modeling of the cytoskeleton in particular.

\section{\label{sec:theory}Theory}

We now consider how the incorporation of cross-linkers with
unfolding immunoglobin
domains will affect the  equilibrium states of a random semiflexible
filament network.  The elasticity of the
\bdchange{unfolding cross-linkers} is highly nonlinear due to the presence of the \bdchange{many}
identical folded protein domains \bdchange{(24 in the case of Filamin)}. It is observed at
large tensile forces ($\sim 100$pN) that
one of these domains will
open, increasing the contour length of the molecule
by about $\sim 20$nm and thereby relaxing
the stress in the system.

As a simple approximation, we model the force extension curve of a
single cross-linker as a sawtooth. Each branch of the sawtooth
represents the entropic elasticity of a cross-linker with a fixed
number of unfolded domains. For simplicity, we take the additional
contour length generated during any unfolding event $\lf$ to be a
constant and the force--extension relation on each branch of the
sawtooth to be linear with spring constant $k_f$. Thus the maximum
force on each branch of the sawtooth, $k_f \lf$, corresponds to the
critical unfolding force of a domain. We also neglect the
rate-dependence of this unfolding force\cite{evans:97} that is found
in nonequilibrium unfolding dynamics.  We refer to this simplified
mechanical model of the physiological cross-linkers as a sawtooth
cross-linker.  Clearly any biopolymer cross-linker that has multiple
unfolding domains that are accessible at physiologically relevant
tensions will act in this manner.  We expect the results presented
below to apply to a network cross-linked by any molecule of this
class.

We now consider the effects of such sawtooth cross-linkers on the zero
frequency mechanics of an elastic network cross-linked by such
agents. In the following we show that the considerations of force
balance alone allow us to relate the distribution of sawtooth
cross-linker extensions modulo the sawtooth length to the, as yet,
unknown distribution of local network compliances in the randomly
cross-linked network.

Imagine the equilibrium state of an individual pulled sawtooth
cross-linker and the region of network surrounding it both before and
after a single unfolding event. We assume that for the relatively
short sawtooth length $\lf$ found in physiological networks, the
response to one unfolding is in the linear regime of the surrounding
network. We model the surrounding effective medium as a single
harmonic spring with spring constant $k$ sampled from some statistical
distribution $\kdist(k)$ which encapsulates the differing local
environments around each cross-link.  Reflecting the network
structure, the cross-linker is connected in series with the effective
network spring. We first set the total strain on the two springs in
series (by fixing their combined length) so that the cross-link is
poised at the maximum extension of its current sawtooth branch.  Then
we consider an infinitesimal increase in the total extension which
drives the unfolding of one more protein domains.

Before the unfolding event the two springs are in force balance at the
top of the previous sawtooth branch, so that $k_f \ell_f = k x$ where
$x$ represents the extension of the medium spring.  After the
unfolding, the system achieves force balance on the next branch of the
sawtooth force-extension curve so that the extension of the
cross-linker is now increased by $\ell_f - d$ while the extension of
the medium spring is decreased to $x - (\ell_f - d)$. Force balance
requires that $d$, the distance between the current extension of the
cross-linker and the edge of the next sawtooth, is given by $d(k) = k
\ell_f/(k + k_f)$. In other words, the combined system once
equilibrated with the cross-linker at its maximal force is now
equilibrated with that cross-linker on its next sawtooth branch at a
smaller force. The strain in the surrounding medium has also decreased
due to the extension of one more protein domain.

To maintain force balance, the cross-linker cannot relax its
length more than $\ell_f - d$. Upon further extension the cross-linker
will only stretch more until another domain unbinds. Thus in
steady-state the sawtooth cross-linker will evenly sample all extensions
(modulo $\ell_f$) between $\ell_f - d(k)$ and $\ell_f$. For a given
value of the spring constant of the medium we expect that the
extensions (modulo $\ell_f$) of the sawtooth cross-linkers $x_f$ to
be uniformly distributed between the bounds given above. This
distribution can be written as
\begin{equation}
P\left(x_f,k \right)=\frac{1}{d(k)} \Theta \left(x_f - \left[\lf -
d(k) \right]\right),
\end{equation}
where $\Theta$ is a step-function.  Integrating over the spring
constant distribution $\kdist(k)$, we write the probability of finding
a given cross-linker length (modulo $\lf$) $x_f$:
\begin{equation}
P\left(x_f\right)=\int_{k_f \frac{\lf-x_f}{x_f}}^\infty \frac{\keff +
k_f}{\lf \keff} \kdist \left(\keff\right) d \keff. \label{eq:pxint}
\end{equation}
The step function fixes the lower limit on the $k$-integral.

There are several important observations to make concerning
Eq.~\ref{eq:pxint}. First we have treated the unfolding events as
being uncorrelated. In this mean field model we have ignored the
effect of additional unfolding events in the surrounding medium. This
simplification allows us to treat the surrounding network as a linear
elastic element with an undetermined spring constant $k$. We will test
this assumption by comparing the predicted relation,
Eq.~\ref{eq:pxint}, to our numerical data.  From this equation we note
that the appearance of an exponential peak in the extension
distribution $P\left(x_f,k \right)$ requires a nontrivial structure
for the distribution of local spring constants $\kdist \left(k
\right)$.  This distribution itself depends on the random connectivity
and material distribution of the network. Below we will examine this
distribution in some detail numerically and we defer speculations
about its form for section \ref{cross-link-ext-stat}, where we discuss
our numerical results.

A second feature of Eq.~\ref{eq:pxint} is that it can be rescaled by
$\lf^{-1}$ in such a way that the dimensionless length distribution
$P\left(x_f/\lf\right)$ is completely independent of the sawtooth
length $\lf$. Thus the distribution $P\left(x_f/\lf\right)$ modulo $1$
depends only on $k_f$ and $\kdist(k)$.  We now turn to a discussion of
the numerical model we used to study random
networks \bdchange{with unfolding cross-links.}

\section{\label{sec:numerics}Numerical Model}

\begin{figure*}
\centerline{\includegraphics{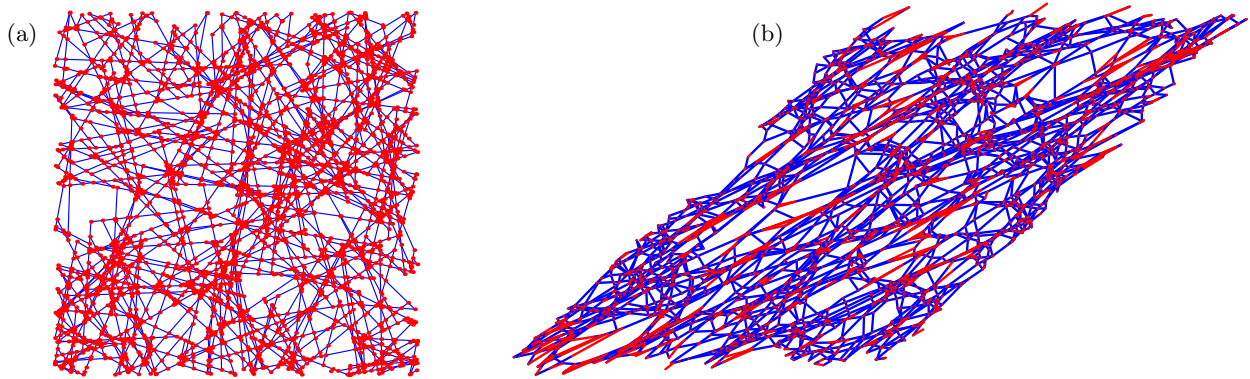}}
\caption{ (color online) Model network showing the F-actin filaments in blue and the
sawtooth cross-linking agents in red. (a) unstrained. (b) $100 \%$ shear.}
\label{fig:network}
\label{fig:actfilsprings}
\end{figure*}

We employ a simple simulational model of a statistically homogeneous,
isotropic network of semiflexible filaments in two-dimensions. These
networks are formed in a manner identical to that of Head {\em et
al.}\cite{head:03a,head:03b,head:03c}, {\em i.e.}  by placing straight
rods of a fixed length at random positions and orientations in a box
with periodic boundaries.  At points where two rods intersect a node
is added to each rod and a cross-linker is added connecting the
nodes. The cross-linkers have zero rest length.  Rods are added until
the average number of cross-links per rod reaches a target value that
we use to parameterize the network density. A model network
constructed by the procedure described above is shown in
Figure~\ref{fig:network}. The sections of rod between nodes are
modeled as linear springs with fixed elastic constant {\em per unit
length} of rod. The cross-linkers exert no constraint torques at the
nodes so that the rods are free to bend there with no energy cost. We
write the Hamiltonian for each filament as
\begin{equation}
{\cal H}=\frac{1}{2}\mu\int \left( \frac{{\rm d}l(s)}{{\rm d}s}
\right)^{2} \,{\rm d}s\,, \label{e:H_Hooke}
\end{equation}
where $dl(s)/ds$ gives the strain or relative change in local
contour length, and $\mu$ is the Young's modulus of the filament
(essentially a spring constant normalized to 1/[length]).

\bdchange{We found that all numerical minimization routines
we tried failed to converge in the non-linear large shear regime
when the energetics of filament
bending were added to our simulational model. This is due to the large
distribution of length scales in the random network, and the resulting
poorly conditioned jacobian. At the same time,
the nonlinear behavior of semiflexible networks
with freely rotating cross-links has been shown to be dominated by
semiflexible filament stretching instead of bending~\cite{onck:05}.
We therefore omit bending energies from our model. This omission allows for 
significant gains in computational efficiency and leads to negligible errors in 
the high-strain state, on which we focus. The
effect of this approximation on our results will be discussed
where appropriate.}

Similarly we anticipate that the results derived here are
essentially independent of network dimensionality since network
connectivity, not the dimensionality of the space in which the
network is embedded, should control the collective mechanical
properties of the system.

Based on the justification given in Section~\ref{sec:theory}, we
model the \bdchange{unfolding} cross-linkers as non-linear springs
with a sawtooth force extension law. Once again, the sawtooth
branches have linear spring constant $k_f$ and length $\lf$.
Though the physiological filament cross-linkers
have a finite number of unfolding domains ($24$), we will let our
simulated sawtooth force extension curve  have an infinite number of
branches.

The network is sheared using
Lees-Edwards boundary conditions (by adding a
constant horizontal offset to filaments that
crossed the top and bottom
boundaries) as
shown in Figure~\ref{fig:network}(b). At
the beginning of each strain step all nodes are moved affinely,
then the node positions are relaxed
through a conjugate gradient routine to a point of local force
equilibrium.
Since the cross-linker force extension curve is a sawtooth, there
are many possible equilibrium states of the network.
We wish to consider the adiabatic, history dependent states of a
strained network, which would in principle require us to use strain
steps resulting in displacements smaller than the sawtooth length $\lf$
so that mechanical equilibrium is achieved in the earliest possible branch.
To save computational time we use a two
step equilibration procedure, which finds a state
close to adiabatic state, but allows for large strain steps.
In the first equilibration step, we replace the sawtooth force law
for all cross-linkers by the following force law:
\begin{equation}
{\bf f} =
\begin{cases}
k_f \bfx & \left|\bfx\right| < \lf, \\
k_f \lf  & \left|\bfx\right| \ge \lf.
\end{cases}
\label{eq:flatforce}
\end{equation}
Thus, beyond the first sawtooth length, a cross-linking
molecule exerts a constant contractile force of magnitude $k_f \lf$.
The combined network of linear elastic rods and constant force
cross-links is equilibrated. Finally, we reimpose a sawtooth force
law for the cross-linkers and equilibrated the network a second
time. Assuming all cross-links act independently as the network relaxes
during this final equilibration step, the force on the cross-link
must be less than $k_f \lf$, so the cross-link will stay on the
same sawtooth branch. Since
the rest of the network was originally equilibrated at the critical
pulling force, the sawtooth force law could not have reached force
equilibrium on any earlier sawtooth branch.
In practice, collective
relaxations of the network push individual cross-links onto
different sawtooth branches in this final step. However, we found
that for a variety of strain step sizes the quantitative behavior
discussed in the results section was identical. \bdchange{This numerical
technique reduced
the required computational time by a factor of ten or more,
allowing us to simulate up to six realizations of random
networks, each with an average of 1100 filaments and 16500 crosslinks,
for every set of material values in this study.
With this technique, we were able to
arrive at statistically meaningful results in a reasonable number of
computer processor hours.}

\section{\label{sec:simresults}Simulation Results}

\begin{table}

\renewcommand{\arraystretch}{1.25}

\begin{tabular}{|c|c|c|c|}
\hline
parameter & symbol & sim units & scaled units \\\hline\hline
Ig domain length & $\ell_f$ & $6.5 \times 10^{-4} $ & $20 \ nm$ \\
cross-link separation & $\ell_c$ & $6.6 \times 10^{-3} $ & $0.2 \ \mu m$ \\
filament length & $\ell_R$ & 0.2 & $6 \ \mu m$ \\
spring ratio & $k_f/\left<k_R\right>$ & $0.06$ -- $66$ & $0.4$ -- $7$ \\\hline
\end{tabular}

\caption{Scaled values of simulation parameters.
The scaled Ig domain length and cross-link separation were taken
from the literature~\cite{freykroy:98,furuike:01}, and were the basis
of the chosen simulational values. The total filament length derives
from the cross-link separation and the decision to use 30 cross-links per
filament on average. This value is consistent with physiological values.
The simulational value of spring ratios were chosen to be broader than
than the typical physiological ratios, so as to capture all possible
regimes.}
\label{tab:values}
\end{table}

The numerical values for all parameters in the simulation, as
well as their corresponding physical values,
are presented in Table~\ref{tab:values}.
We present data for networks with a filament density such that there
are on average $30$ cross-links per filament. The filaments have a unique
length of $\ell_R=0.2$ measured in units of the length of the
unstrained, square simulation box. This gives an average cross-link
separation of $\ell_c \approx 6.6 \times 10^{-3}$ in simulation units.
For these values we find
negligible system-size effects. The other
length scale in the system is $\lf$, the length
of the sawtooth in our approximate form of the cross-linker
force-extension relation. We found that for $\ell_R/\ell_c\gtrsim
15$ and $\lf/\ell_c$ constant, the network behavior is independent
of $\ell_R$.
The distance between cross-links in physiological F-actin
networks has been quoted as
$\ell_c = 0.2 \  \mu m$~\cite{freykroy:98}. For Filamin cross-links
each unfolding domain
adds $\lf=21 \ nm$ of length~\cite{furuike:01}. Thus the physiological
ratio of sawtooth length to cross-link separation is
$\lf/\ell_c = 0.1$. We expect this ratio to be similar for all
domain-unfolding cross-linkers. To explore the dependence of our results
on this ratio and thus on the particular type of sawtooth cross-linker, we
present data for $\lf/\ell_c \approx 0.1$ and $\lf/\ell_c \approx 0.02$
(respectively, we used $\lf=6.5 \times 10^{-4}$
and $\lf=1.3 \times 10^{-4}$ in simulation units).

There are two energy scales in the system determined by the
extensional modulus of the filaments $\mu$ and the spring constant
of the cross-linkers. To fix the energy scale in the problem we set
the extensional modulus of the filaments to unity. The average
spring constant for a filament segment can then be determined from
the mean distance between cross-links, or, in other words, the
network density: $\left<k_R\right> = 1 / \left<\ell_c\right> = 150$.
Physically, actin is a worm-like polymer, so the effective spring
constant of each free length of polymer should be
$k_R \sim k_B T \ell_p^2 / \ell_c^4$~\cite{mackintosh:95}, where
$\ell_p$ is the persistence length of F-actin, which is approximately
$17 \ \mu m$~\cite{freykroy:98}. Thus at room temperature,
a $0.1 \ \mu m$ segment of actin will have
$k_R \sim 12 \ pN / nm$, while a
$0.2 \ \mu m$ segment will have $k_R \sim 0.7 \ pN / nm$. Because of
the  variability in this effective spring constant as
a function of $\ell_c$, we choose to simulate a range of actin
to \bdchange{cross-linker} spring constant ratios.
The  measured unfolding force for an Ig domain is
around $150 \ pN$~\cite{furuike:01} -- treating the cross-linker
as a linear spring between unfolding events and using the
sawtooth length given above, we arrive at a
physiological spring constant of $k_f \sim 8 \ pN / nm$.
Thus the relevant dimensionless ratio of spring constants
to study for physiological systems may lie in the
range between $0.4 < k_f/\left<k_R\right> < 7$.
In simulation units, the range of cross-linker
spring constant values reported here was between $10^1 <
k_f < 10^4$, which corresponds to a range of spring constant
ratios between  $0.06 < k_f/\left<k_R\right> < 66$.

\subsection{Elastic moduli}

\begin{figure*}
\centerline{\includegraphics{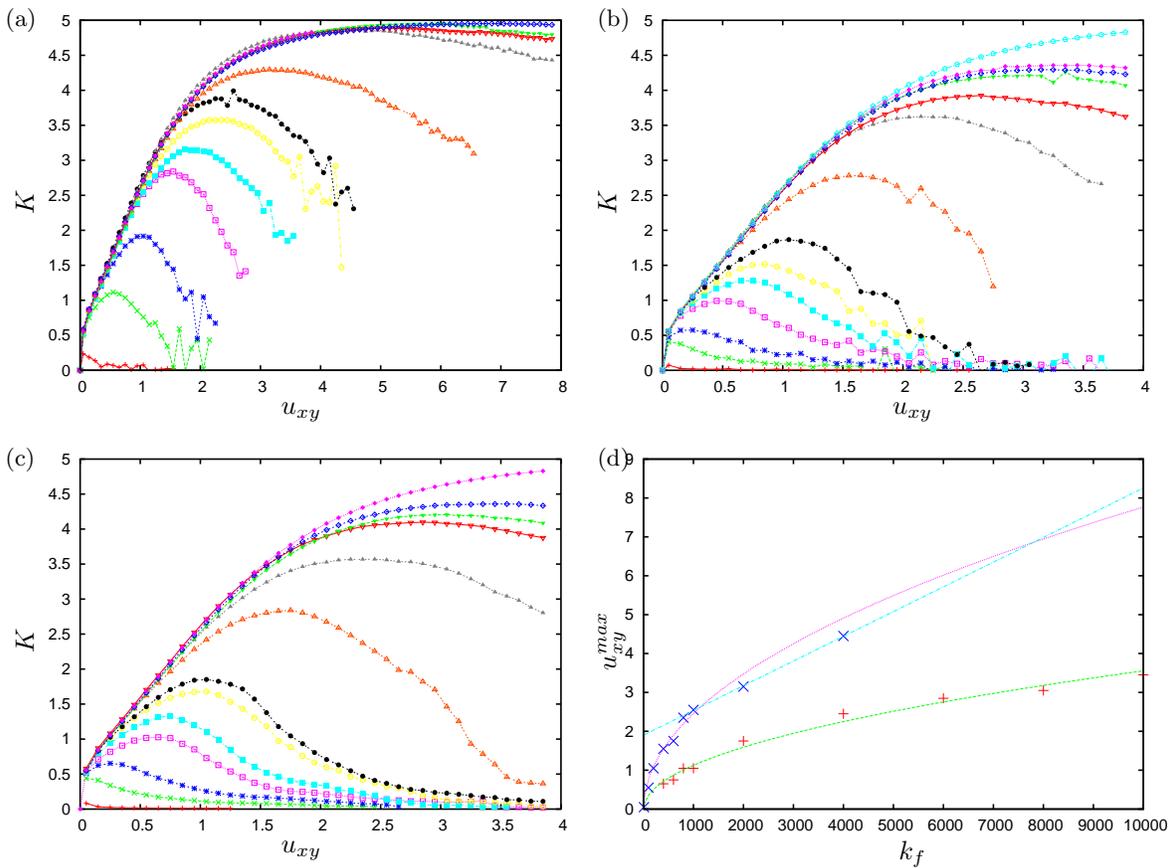}}
\caption{(color online) Differential moduli versus strain for several different
cross-linker spring constants $k_f$. Modulus $K$ is given in simulation units.
(a) Shear modulus versus shear strain for $\lf/\ell_c = 0.1$.
(b) Shear modulus versus shear strain for $\lf/\ell_c = 0.02$.
(c) Shear modulus versus shear strain, for constant
cross-linker force $f=k_f \lf$ with $\lf/\ell_c = 0.02$.
From lowest to highest curves in each graph, the \bdchange{cross-linker} spring constants are respectively
$10$, $100$, $200$, $400$, $600$, $800$,
$1000$, $2000$, $4000$, $6000$, $8000$, $10000$ and inextensible cross-links.
(From  Section~\ref{sec:simresults}, the average
filament segment spring constant in these
units is $\left<k_R\right> \approx 150$.) (d) Strain value $u^{max}_{xy}$
corresponding to the peak differential
modulus from the data in (c) (lower points) and (a) (upper points)
as a function of $k_f$. The curved lines
are fits to $\sqrt{k_f}$. The straight line is a fit to $ax+b$ for
values of $k_f \ge 1000$ and $\lf/\ell_c = 0.1$.
}
\label{fig:modvstrain}
\end{figure*}

Since we have omitted the bending rigidity of the filaments,
these networks necessarily have vanishing elastic moduli in their
unstrained state\cite{kellomaki:96}. \bdchange{It has been
shown that {\it in vitro} F-actin Filamin networks do have a finite modulus
at zero shear, though the differential modulus increases by a factor
of $100$ or more upon shearing~\cite{gardel:06}.}
Our networks quickly
develop non-zero moduli under   finite strain. To
characterize the nonlinear mechanical properties of the network we
 measure the differential moduli $\partial \sigma_{ij}/\partial
u_{ij}$ as a function of applied strain.
Fig.~\ref{fig:modvstrain} shows how the differential shear
modulus evolves as a function of strain.  As expected, the differential
moduli all vanish for zero applied strain. Upon increasing applied
strain, they grow monotonically to some maximum where the network is
stiffening most dramatically under further strain increments. At
still larger values of the applied strain, the stress
plateaus so that the differential moduli shown here decay to zero.
In part (a) of
Fig.~\ref{fig:modvstrain} we show the differential shear
modulus $\shrmod(u_{xy})$ for networks with $\lf/\ell_c = 0.1$
as computed from the additional shear
extension of the network already under shear $u_{xy}$. In part
(b) of this figure we show the differential
shear modulus $\shrmod(u_{xy})$ for networks with $\lf/\ell_c = 0.02$.
Though not shown, we also measure the uniaxial
extension modulus $\strmod$
for the case $\lf/\ell_c = 0.02$. From a
comparison of the two measurements we determine the effective Poisson
ratio $\nu$ as function of applied strain\cite{landau:vol7} using
\begin{equation}
\label{Poisson}
\nu = \frac{1}{2} \frac{3 \strmod - 2 \shrmod}{3 \strmod + \shrmod}.
\end{equation}
For intermediate deformations, where the network is strain stiffening,
the scale of $\strmod$ is about four
times that of $\shrmod$, so $\nu \approx 0.38$.

The peak in the differential modulus was observed to
coincide with the appearance of ``tears'' in the network.
These tears were actually
clusters of highly extended cross-linkers.
Eventually, a single cluster
of highly extended cross-linkers  percolates across the
network. \bdchange{Thus, the modulus peak occurs when  a significant
fraction of the cross-linkers are at their unfolding threshold. }

\bdchange{The formation of the fragile state characterized by cross-linker unfolding requires a
high-stress state. Since F-actin is found to rupture under tensile loads of 600pN~\cite{tsuda:1996},
one may ask whether F-actin rupture precludes the 
formation of the fragile state characterized by cross-linker unfolding. 
To check this we measured the tension 
along the actin filaments at the onset of cross-linker unfolding; we found that,
although there are multiple cross-linkers connected to each filament,
the force due to adjacent cross-linkers tend to cancel, so that
stresses do not accumulate along the actin filaments.
Using a conservatively high estimate unfolding force of Filamin, 150 pN, we find that,
at the onset of cross-linker unfolding, the stresses on
individual segments of actin filament (between cross-links)
were all below $4$ times the cross-linker unfolding tension (600pN), and
$90\%$ of filaments segments were at tensions below $3$ times
that value (450pN).  At larger strains these forces increase, and
at strains $1.5$ times the differential modulus peak, up to $20\%$
of the actin filaments may be under more than 600pN of tension,
effectively breaking those strands. The breakdown of the network
at large strains is safely above the
region of interest in this work making the formation of the fragile state possible. 
Finally, we note that the unfolding tension of the cross-linker is actually an out-of-equilibrium 
quantity. At slower rates of tensile loading the cross-linkers will unfold typically at lower tensions
due to thermally activated processes. Such thermal effects at lower loading rates, 
serve to widen the range of strain states
in which cross-linker unfolding occurs before F-actin failure. 
We will discuss the scaled tensions in the network
further in Section~\ref{sec:discussion}.
}

\bdchange{Since the tearing of the network corresponds to a segregation
into actin-rich and cross-linker rich regions,}
we can thus approximate the
compliance of the network as the compliance of a composite system.
One part of the composite is the filament dominated parts of the
network in which the cross-linkers are not greatly extended and the
other part is the region of large cross-linker stretch. The
effective modulus of the composite system can then be approximated
as two nonlinear springs in series.

The force law for the filament dominated parts of the
network can be inferred from the top curve
in Fig.~\ref{fig:modvstrain}(b), which shows the differential
shear modulus for networks with non-compliant cross-links. The
shear modulus increases steadily at low shear as the individual
filaments align with the shear direction. Above $200\%$ strain,
the modulus starts to plateau. In the limit of complete shear
alignment, the extension of individual filaments is
linear in the shear and thus the differential modulus is constant, with
value $\shrmod \left(u_{xy}\right) = n_{ave} k_R$, where $n_{ave}$
is the average number of filaments per unit cross sectional length.
At low and intermediate values of the shear, the dependence
of the modulus on absolute shear is complicated, reflecting
simultaneous filament extension and alignment. In practice, we
observe the growth of the modulus in this regime to be approximately
linear in the shear.

In contrast, the part of the network dominated by highly extended
cross-links can only sustain a finite maximum stress per unit area,
since each cross-linker can only exert a maximum force of $k_f \lf$.
In the limit of complete shear alignment, the maximum stress is
given by
\begin{equation}
\sigma_{max} = n_{cr} k_f \lf,
\label{eq:sigmax}
\end{equation}
where $n_{cr}$ is the number of
cross-linkers per unit of cross-sectional length that take part in the
crazing, or splitting, of the network. $n_{cr}$ is determined by the
shortest percolation path across the network. We surmise that the
differential modulus of the composite system
reaches it maximum value at the onset of
crazing.
The expression above for the maximum stress implies that the onset of
crazing is determined purely by the product $k_f \lf$. Indeed,
Figure~\ref{fig:modvstrain}(c) shows the modulus for sheared networks
where the cross-linker force law is simply that given in
Eq.~\ref{eq:flatforce}: essentially a constant force cross-link
with force $k_f \lf$.
The curve is nearly identical to Figure~\ref{fig:modvstrain}(b),
for the sawtooth force law. A constant force law is
by definition independent of network extension, so the differential modulus
$\partial \sigma_{ij}/\partial u_{ij}$ is zero in a stretched,
cross-linker dominated region of the network which has plateaued at the maximum
stress $\sigma_{max}$.

We now consider shear strain that maximizes the differential shear
modulus.  The maximum differential shear modulus occurs at the
crossover point, $u_{xy}^{max}$, when the applied stress approaches
$n_{cr} k_f \lf$ as shown in Eq.~\ref{eq:sigmax}.  For smaller
$k_f/k_R$, the crossover occurs in the linear growth region of the
differential modulus, where $\sigma_{xy} \sim n_{ave} k_R u_{xy}^2$,
so the shear at maximum scales as $u_{xy}^{max} \sim \sqrt{k_f/k_R}$.
For larger $k_f/k_R$ the crossover occurs in the region of constant
differential shear modulus, so the shear at maximum scales as
$u_{xy}^{max} \sim k_f/k_R$. The plots in
Figure~\ref{fig:modvstrain}(d) shows $u_{xy}^{max}$ versus $k_f$ for
the data in Figure~\ref{fig:modvstrain}(a) and (b).  The data for
$\lf/\ell_c = 0.02$ is consistent with $u_{xy}^{max} \sim \sqrt{k_f}$.
For $\lf/\ell_c = 0.1$, the strain at maximum scales as $u_{xy}^{max}
\sim \sqrt{k_f}$ for $k_f\le1000$, then approaches a linear dependence
on $k_f$ at higher $k_f$.

For larger relative values of $\lf$ the
splitting of the network, and therefore the drop-off of the
differential moduli, is suppressed at lower shear values. It does
not occur until a finite fraction of the cross-linker population
is stretched beyond its
initial sawtooth length $\lf$. Still, in the limit of
large strain, the modulus
is mainly determined by the combination $F_{max} = k_f \lf$.
In the
next section we will show that most stretched cross-linkers reach
equilibrium near $F=F_{max}$, so it is natural that the network
elastic response is essentially that of  a network
with constant force cross-linkers.

In living cells, the action of molecular motors
might lower the threshold of cross-link extension beyond
the sawtooth length by prestressing the network. Indeed, it has been
shown~\cite{gardel:06} that prestress is necessary to
replicate physiological
conditions in {\em in vitro} F-actin Filamin systems.
Physiologically, unfolding-based behavior should manifest itself once the
average stress per filament exceeds the Ig domain
unfolding stress of $\approx 150$ pN~\cite{furuike:01}.
For a dense network with $10$ filaments per
$\mu m^2$ of cross section the critical prestress
would amount to $1.5$ kPa, while for sparse network with on the
order of  $1$ filament per $\mu m^2$ the critical prestress would be
$150$ Pa or less.

\begin{figure*}
\centerline{\includegraphics{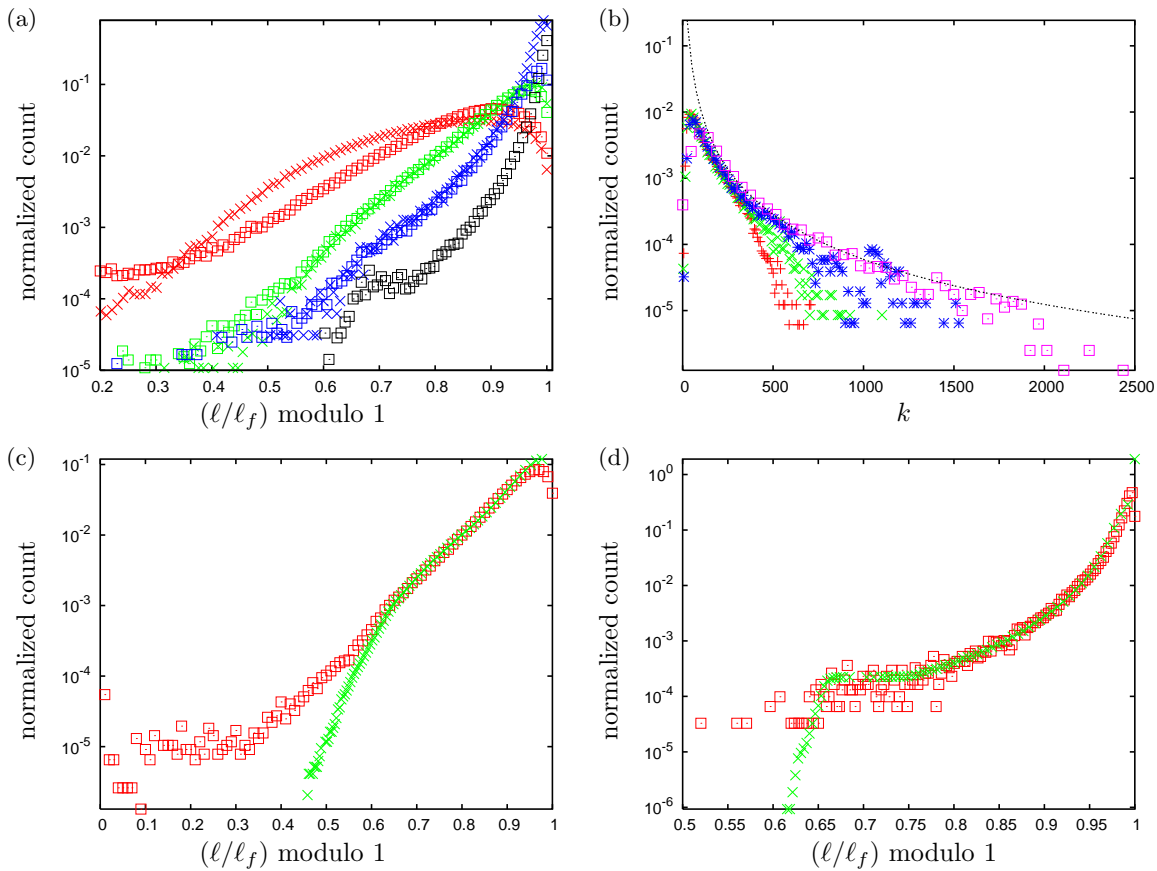}}
\caption{ (color online) (a) Distribution of normalized
cross-linker lengths $\ell/\ell_f$ modulo $1$ in equilibrated networks with,
from shallowest to steepest slopes respectively, $k_f=100$,
$600$, $2000$, and $6000$ (where $\left<k_R\right>=150$).
For the three smaller values of $k_f$,
values were given for $\lf/\ell_c = 0.1$ (crosses) and for
$\lf/\ell_c = 0.02$ (boxes). For $k_f=6000$ no significant pileup was
measured for $\lf/\ell_c = 0.1$. In plots (b) -- (d) we only
plot values for
$\lf/\ell_c = 0.02$. (b) Measured effective medium spring
constant distributions for $k_f=600$, $1000$, $2000$ and $6000$ (higher
$k_f$ have longer tails). The dashed line shows $k^{-2.25}$.
(c) Distribution of normalized
cross-linker lengths $\ell/\ell_f$ modulo $1$ for $k_f=600$ (boxes)
compared to distribution predicted by Eq.~\ref{eq:pxint} for the
data in (b) (crosses).
(d) Distribution of normalized
cross-linker lengths $\ell/\ell_f$ modulo $1$ for $k_f=6000$ (boxes)
compared to distribution predicted by Eq.~\ref{eq:pxint} for the
data in (b) (crosses).} \label{fig:modulus} \label{fig:pileup}
\end{figure*}

\subsection{Cross-linker extension statistics}
\label{cross-link-ext-stat}

The crazing or splitting of the network discussed in the last
section occurs once the local strain
forces experienced by cross-links exceeds $k_f \lf$, the maximum
force sustainable by a cross-link sawtooth branch. We found that, in
the subset of all cross-links which had been extended through at
least one ``unfolding event'' (i.e., those cross-links which had changed
sawtooth branches at least once), a characteristic distribution
of equilibrium lengths (modulo $\lf$) emerges that is
independent of total strain.
Figure~\ref{fig:pileup}(a) shows the measured equilibrium distributions
of cross-link lengths, modulo the sawtooth length $\lf$, for a
representative set of strained networks with $\lf/\ell_c = 0.1$ or
$\lf/\ell_c = 0.02$
and several values of spring constant $k_f$.
For values of  $k_f / \left<k_R\right> <10$ the statistical
weight for finding a cross-link extension (modulo $\lf$) appears
exponentially enhanced towards length $\lf$ where the domains
unbind. For values $k_f / \left<k_R\right> >10$ the statistical
weight for finding a cross-link extension (modulo $\lf$)
grows faster than exponentially near length $\lf$.
For $k_f / \left<k_R\right> \ge 4$ the
distributions of cross-link lengths were
identical for both measured values of $\lf$. At lower $k_f$ there
were significant differences between the $\lf/\ell_c = 0.1$ and
$\lf/\ell_c = 0.02$ data, with the former showing a less pronounced
pileup towards length $\lf$. The source of this deviation could be
non-linearities in the response of the softer networks over the
longer length scale of $\lf/\ell_c = 0.1$ -- our effective
spring model assumed linear response.

According to the model developed in Section~\ref{sec:theory}, the
cross-link length distribution is determined completely by the
distribution of local effective network spring constants.
In particular, there is nothing
in Eq.~\ref{eq:pxint} that
would create an exponential in the length distribution
unless that exponential were already
contained in $\kdist(k)$, the effective spring constant
distribution function.

Figure~\ref{fig:modulus}(b) shows
the measured distribution of effective network spring constants
experienced by the cross-linkers sampled in
Figure~\ref{fig:pileup}(a). This distribution was found by probing
one cross-link at a time, changing the force on the cross-link and
numerically re-minimizing over lattice displacements to find the
corresponding change in length.
We find that there is indeed a non-trivial
and rapidly decaying distribution of effective spring constants.
For high $k_f$ the spring constant distribution
approaches a power-law with exponent $-2.25$, while for lower
$k_f$ the distribution is truncated at high effective $k$,
becoming nearly exponential.

We may use Eq.~\ref{eq:pxint} to calculate $P(x_f)$, the
expected distribution of cross-linker lengths modulo $\lf$,
for the cases of exponential or power law $\kdist(k)$.
For the the power law case with $\kdist(k)\sim k^{-p}$,
Eq.~\ref{eq:pxint}
yields a cross-link length distribution
of the form
\begin{multline}
\label{pow-dis}
P(x_f) \simeq \frac{k_f^{1-p}}{p(p-1)\lf}
\left[(p-1)\left(\frac{x_f}{\lf-x_f}\right)^p + \right. \\
\left. p \left(\frac{x_f}{\lf-x_f}\right)^{p-1} \right].
\end{multline}
This form for $P(x_f)$ diverges at $\lf$ for any $p>0$.
Alternately, if the effective spring constants were exponentially
distributed,
with form $\kdist(k) \sim e^{-k/\bar{k}}$,
then $P(x_f)$ would take the form
\begin{multline}
\label{exp-dis} P(x_f)\simeq \frac{\bar{k}}{\lf}
\exp\left(\frac{k_f (x_f - \lf)}{\bar{k} x_f}\right) + \\
\frac{k_f}{\lf} \Gamma
\left(0,\frac{k_f (\lf - x_f)}{\bar{k} x_f}\right),
\end{multline}
where the constant $\bar{k}$ is an undetermined material parameter and $\Gamma$
is the incomplete Gamma function.
This expression grows exponentially near $x_f = \lf$.

Eqs.~\ref{pow-dis} and \ref{exp-dis}, together
with the data in Figure~\ref{fig:modulus}(b),
are consistent with our
observation that the cross-linker length distribution
grows exponentially for low $k_f$ and even faster for
high $k_f$, since the respective forms for $\kdist(k)$
for these two cases are exponential and power law.
Figure~\ref{fig:modulus}(c) shows the result of
numerically evaluating Eq.~\ref{eq:pxint} on the
data from Figure~\ref{fig:modulus}(b) for $k_f / \left<k_R\right> =4$. The
agreement between the theoretical prediction for
$P(x_f)$ and the actual measured distribution is very
good at high $x_f$. Similarly,
Figure~\ref{fig:modulus}(d) shows the result of
numerically evaluating Eq.~\ref{eq:pxint}
for the measured $\kdist(k)$ for $k_f / \left<k_R\right> =40$.
The agreement between theoretical prediction
and measurement is outstanding.

\begin{figure}
\centerline{\includegraphics{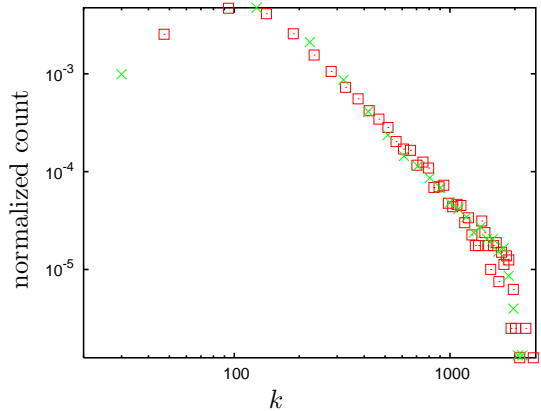}}
\caption{ (color online) Measured effective medium spring
constant distributions for $k_f / \left<k_R\right> =40$.
Box points are for
a network with sawtooth cross-linkers, while the cross points
are for a network with linear spring cross-links.
Plots for
$\lf/\ell_c = 0.02$. } \label{fig:modulus2}
\end{figure}

Figure~\ref{fig:modulus2} again shows $\kdist(k)$
for $k_f / \left<k_R\right> =40$.
The distribution function for sawtooth force law cross-linkers
is compared to the distribution for a network which was
created and equilibrated under strain with linear force law
cross-linkers. Both distributions are identical, demonstrating that
the presence of the sawtooth force law does not determine the
distribution of $\kdist(k)$. This justifies our use of a mean-field
approach in Section~\ref{sec:theory}, since $\kdist(k)$ is not
influenced by $P(x_f)$. The further implication is that
the origin of the measured $\kdist(k)$ is purely geometrical.

We have seen that for large values of the sawtooth spring constant
$k_f$ the local spring constant distribution in the random network has a power
law decay in the large $k$ regime as shown by the boxed points in Fig.~\ref{fig:modulus}(b).
These points are fit by the dashed line representing a power law with exponent $-9/4$.
The power law behavior is also clearly visible in the log-log
plot in Fig.~\ref{fig:modulus2}, which
shows the higher $k_f$ data with $\ell_f/\ell_c = 0.02$.
For smaller values of $k_f$, {\em e.g.}
the crosses in the Fig.~\ref{fig:modulus}(b), the local spring
constant distribution exhibits a clearly exponential decay.
Considering the full set of
the numerical data shown in Fig.~\ref{fig:modulus}(b) for varying values of $k_f$
ranging (in simulation units) from 600 to 6000, we note the following trends. First, the
distribution appears to transition from a power law form at higher $k$ in the case of stiffer
cross-linker springs, {\em i.e.} larger
values of $k_f$, to an exponential form for softer cross-linker springs.
Secondly, the maximum values of local spring constants observed in the system appear
to be generically smaller than the cross-linker spring
constant for that system. Finally, each data set
for a given $k_f$ appears to follow the power law decay for smaller values of $k$ before reaching
an exponentially decaying terminal regime at the largest measured values of $k$.

We now speculate as to the origin of these phenomena beginning with an examination of the
power lay decay seen for smaller values of $k$ in all the data sets.  Since this effect
is more robust at larger values of the cross-linker spring constants, consider first the
limiting case of perfectly incompliant cross-linkers. Here the local effective
spring constant of the network must be completely determined by the
filaments, and, in particular, by the distribution of filament
lengths between cross-links.  The distances between cross-links,
$\ell_c$, for a random network such as ours are exponentially
distributed, with
\begin{equation}
P(\ell_c) d \ell_c \sim e^{-\ell_c /\left<\ell_c\right>} d \ell_c.
\label{eq:kallmesdist}
\end{equation}
as shown by Kallmes and Corte~\cite{kallmes:60}.
Since the spring constant of each filament segment is proportional to
$1/\ell_c$, the probability of the occurrence of a filament segment spring constant
between $k$ and $k + dk$ takes
the form
\begin{equation}
P(k)  dk \sim e^{-1/(\left<\ell_c\right> k)} k^{-2} dk.
\label{eq:k2tail}
\end{equation}

Furthermore, if we assume that softer regions of the network from which the smaller values of
local spring constant $k$ are measured form small pockets in a material that is generally
less compliant, then it is reasonable to suppose that the deformation imposed on the network
due to the external forces applied in those softer regions will lead to a highly localized deformation
involving only a few network springs.  Then the local effective spring constant
around those cross-links is
determined primarily by a few filament segments whose
effective spring constants statistics will be governed by Eq.~\ref{eq:k2tail}.
For large values of $k$, this expression approaches $k^{-2}$, which is
consistent with our findings for large $k_f$. The exponential in
Eq.~\ref{eq:k2tail} becomes unimportant
for $1 \lesssim \left<\ell_c\right> k$ so that for $150 \lesssim k$ we
expect to see a only
power law distribution of local spring constants with an exponent of $-2$.

The effective spring constant of any two springs in series is less than the spring constant
of either of them. Since all filaments are connected to one another through
cross-linkers with spring constant $k_f$, it is clear that this power law tail
in Eq~\ref{eq:k2tail} cannot extend to
values of $k$ greater than $k_f$.  If we consider the large-$k$ tail of the spring
constant distribution, we must look at rarely found regions in the network that lie
on chains of anomalously stiff network springs corresponding to very short network filaments.
Each of these chains of springs is
made up of a number of statistically independent springs
connected in series. In order to find an extremely large value of
the effective spring constant $k$ it must be that for one of the
force paths {\em all of the constituent spring constants are large},
since the compliance of the springs in series will be dominated by
any single soft spring. We expect the probability of such a rare
event to be Poisson distributed so that, in the
high-$k$ tail, the distribution $\kdist(k)$ takes the form
\begin{equation}
\label{Poisson-tail} \kdist(k) \sim H(k) e^{-k/\bar{k}}
\end{equation}
where $H$ is some regular function characterizing the small-$k$
behavior of the distribution ($H(x) \rightarrow \mbox{const}$ as $x
\rightarrow \infty$) discussed above and the constant $\bar{k}$ is undetermined
by this heuristic argument.
Such an exponential distribution is consistent with the data for
low $k_f$ presented in Figure~\ref{fig:modulus}(b).

Interestingly, we do indeed see that each distribution has an exponential tail
for values of $k$ approaching an upper limiting value of $k_f$.  The value of
this upper limit appears to be $\propto k_f$ with a coefficient of
proportionality of order unity. This suggests that if the stiff chains of springs associated with
the stiffest region of the material contains of order $n$ springs in series then there must be
of order $n$ such chains in parallel to account for the upper limit.

The principal effect of changing $k_f$ on the elasticity of the random network is to push the
cross-over from a power law decay to the exponential tail to larger values of $k$.  We have
presented arguments for both this lower-$k$ power law and higher-$k$ exponential dependence of
the local spring constant distribution. We cannot account for approximately ten percent
discrepancy between the measured and predicted exponents for the power law and we cannot
estimate using these arguments the cross-over value of $k$ between these two behaviors. Nevertheless
we see regardless of the form of the local spring constant distribution, Eq.~\ref{eq:pxint} accurately relates
that distribution to the cross-linker extension
distribution as shown by Fig.~\ref{fig:modulus}(c) and Fig.~\ref{fig:modulus}(d)

\section{Discussion}
\label{sec:discussion}

In this work we have studied a composite network of linear
elastic elements having randomly distributed spring constants
cross-linked by linkers that have the
sawtooth force-extension relation common to proteins with repeated
unfolding domains.  While this system is clearly an
oversimplification of both the chemical complexity and semiflexible
character of the cytoskeleton, these networks retain one important
micro-architectural feature of the F-actin networks in that forces must
propagate from filament to filament through a linking molecule
exhibiting a highly nonlinear (sawtooth) force response to strain.
We suspect that our model system can thus inform the understanding
of the mechanical properties of physiological cytoskeletal networks,
which are the subject of recent theoretical work and are being
probed experimentally with increasing quantitative
accuracy\cite{lau:03,hoffman:05,deng:06}.

Our most striking result is the observation of the development of a
highly fragile mechanical state in the network at large strains. At
moderate strain, a nontrivial number of cross-linking molecules reach
a critical state where they are poised to unfold another domain. The
presence of fluctuating internal stresses in the cytoskeleton produced
by variations in molecular motor activity and/or thermally generated
fluctuations can act on this highly fragile state to produce strain
fluctuations at all frequency scales, due to the broad distribution of
local energy wells in the system.  Thus, the formation of this
critical state under applied stress may explain a particular feature
of the low-frequency strain fluctuations as observed by intracellular
microrheology.

\bdchange{The inclusion of sawtooth cross-linkers in the analysis of 
the mechanics of semiflexible networks may explain some 
features of the nonlinear elasticity of the cytoskeleton. 
It is well-known that {\em in vitro} F-actin networks when 
cross-linked by rigid, inextensible molecules generically 
show strain hardening\cite{gardel:04} at least at network 
densities consistent with the affine response regime\cite{head:03c}.  
We did not include the nonlinearity of the longitudinal 
compliance of the actin filaments, which accounts for 
this strain-hardening.  The mechanics of unfolding 
cross-linkers presents an additional nonlinear effect that 
decreases the differential modulus. Taking both effects into 
account, we expect that highly strained F-actin/Filamin gels 
to strain harden {\em less than} than that predicted by Ref\cite{mackintosh:95} 
and eventually show a strain softening regime at high strains.

While strain hardening is a hallmark of highly cross-linked 
semiflexible networks, the {\em in vivo} cytoskeleton has 
been seen to strain soften\cite{fredberg:06}.
Based on this work we suggest that strain softening in densely 
cross-linked semiflexible networks may be the result of the 
unfolding of specialized cross-linking proteins 
in it.  From our numerical data it is clear that cross-linking 
of these networks  by unfolding linkers can generate 
a strain softening material at large enough stresses so that 
domain unfolding occurs. It remains to be seen how or whether 
the interaction of these cross-linkers with motor proteins 
shifts this nonlinear elastic regime to lower applied strains.}

We believe that this work suggests the appropriate theoretical
framework for understanding the underlying mechanism by which the
system reaches this highly fragile state. We note, however, that much
remains to be done in order to develop this understanding into a
theory that makes quantitatively accurate predictions.  In addition it
is clear that more precise numerical explorations of the network are
required in order to better characterize both the local elastic
constant distribution in the network as well as the cross-linker
length distribution.  Nevertheless, it appears that the creation of this
fragile state is a robust phenomenon.

In addition, recent experiments by the Janmey group report that
rigidly cross linked semiflexible networks generically show a large,
negative first normal stress coefficient\cite{mackintosh:06}.  We know
of no such measurements on the {\em in vivo} cytoskeleton. Noting that
sawtooth cross-linking of these networks
changes their nonlinear elasticity, we believe it will be important to
study the effect of \bdchange{unfolding cross-linkers} on this nonlinear elastic
response as well.

There are a number of other extensions of this work that remain to be
considered. \bdchange{We are currently working to add thermalized
subcritical unfolding, along with energy input through the action
of simulated molecular motors. The non-equilibrium steady state
thus established will be closer to the conditions inside cells.
The strain rate dependent unfolding force~\cite{evans:97} then
depends on the frequency spectrum of molecular motor activity,
so that the effective unfolding threshold, and thus all other
forces in the network, may be shifted down by a
factor of $2$ or more.}
The development of a more complete model that includes the
effect of internally generated random stresses due to the action of
molecular motors will be an important step towards the direct
calculation of the low-frequency dynamics of this biopolymer gel of
fundamental biological importance.

\section*{Acknowledgments}

BD and AJL thank J.C. Crocker for providing unpublished data and
for enjoyable discussions. BD also thanks David Morse for
enlightening discussions. AJL was supported in part by
NSF-DMR0354113. BD acknowledges the hospitality of the UCLA
department of Chemistry and the California Nanoscience Institute
where part of this work was done. BD also acknowledges partial
support from NSF-DMR0354113 and the Institute for Mathematics and its
Applications with funds provided by the
National Science Foundation.

\end{document}